**Title of the article:**

Agile Software Development Methods: A Comparative Review1

**Authors:**

Pekka Abrahamsson, Nilay Oza, Mikko T. Siponen

**Notes:**

- This is the author's version of the work.
- The definite version was published in: Abrahamsson P., Oza N., Siponen M.T. (2010) Agile Software Development Methods: A Comparative Review1. In: Dingsøyr T., Dybå T., Moe N. (eds) Agile Software Development. Springer, Berlin, Heidelberg
- Copyright owner's version can be accessed at https://link.springer.com/chapter/10.1007/978-3-642-12575-1_3

# Agile Software Development Methods: A Comparative Review[1]

*Pekka Abrahamsson, Nilay Oza and Mikko T. Siponen*

**Abstract**: Although agile software development methods have caught the attention of software engineers and researchers worldwide, scientific research still remains quite scarce. The aim of this study is to order and make sense of the different agile approaches that have been proposed. This comparative review is performed from the standpoint of using the following features as the analytical perspectives: project management support, life-cycle coverage, type of practical guidance, adaptability in actual use, type of research objectives and existence of empirical evidence. The results show that agile software development methods cover, without offering any rationale, different phases of the software development life-cycle and that most of these methods fail to provide adequate project management support. Moreover, quite a few methods continue to offer little concrete guidance on how to use their solutions or how to adapt them in different development situations. Empirical evidence after ten years of application remains quite limited. Based on the results, new directions on agile methods are outlined.

## 3.1 Introduction

Agile – denoting "the quality of being agile; readiness for motion; nimbleness, activity, dexterity in motion" (http://dictionary.oed.com) – software development methods attempt to offer once again an answer to the eager business community asking for lighter weight along with faster and nimbler software development processes. This is especially the case with the rapidly growing and volatile Internet software industry as well as with the emerging mobile application environment. The new agile methods have evoked a substantial amount of literature (e.g., Cockburn 2002; Highsmith 2002a; Martin 2002) and debates (e.g., Yourdon 2000; Highsmith 2001; Highsmith 2002b). While some authors claim (e.g., Baskerville et al. 2002; Meri-Salo et al. 2005) that agile software development methods do not offer anything new with regard to software development principles, agile advocates maintain that the agile principles represent a new – radically different – paradigm in software engineering (cf. Rajlich 2006).

---

[1] An early version of this chapter was presented at International Conference on Software Engineering in 2003 (ICSE 25).

Despite of widespread application of agile methods in wide range of different industrial contexts, there is still no clear agreement of what are the focal aspects of agile methods. Some claim that these are simplicity and speed (Beck 1999; Highsmith and Cockburn 2001; McCauley 2001) while others suggest them to be collaboration, co-ordination and communication (REF missing). In development work, accordingly, development groups concentrate only on the functions needed at a given moment, delivering them fast, collecting feedback and reacting rapidly to changes in business and technology (Aoyama 1998; Fowler and Highsmith 2001; Müller and Tichy 2001; Boehm 2002). Since 1998, at least twelve slightly varying definitions of what is agility in software development have been offered (Kettunen 2009). The apparent conceptual confusion has not slowed down the adoption of agile methods in industrial settings. Anecdotal evidence has grown substantially in the 2000's as well as the number of experience reports in conferences and other business driven events.

The number of new agile methods increased rapidly during the first years and this phenomenon is not showing any signs of fading still. This has resulted in a situation where researchers and practitioners are not aware of all the available approaches or of their suitability for varying real-life software development situations. As for researchers and method developers, the lack of unifying research hinders their ability to establish a reliable and cumulative research tradition.

The aim of this study is to order and make sense of the different agile approaches that have been proposed. The result of this analysis will make practitioners better aware of the available agile methods, who will thus be in a better position to understand the features of agile methods, and hence to choose the most appropriate method in a more informed way.

An analytic framework is constructed for scrutinizing and guiding the review of the existing agile methods. The framework encompasses six perspectives: project management support, software development life-cycle coverage, availability of concrete guidance for application, adaptability in actual use, research objective, and empirical evidence. The results show that the existing methods cover various phases of the life-cycle, without, however, offering any rationale. The majority of them do not present adequate support for project management, and only few methods offer concrete guidance to support their solutions or adaptations in different development situations. Furthermore, while method developers' research objective is predominantly means-end oriented (technical), related empirical evidence to date remains very limited. Based on the results, new research directions on agile methods are outlined.

The rest of the chapter is composed as follows. The second section presents a short overview of the existing agile methods. The third section presents the analytical perspectives and the rationale for them. The fourth section presents a comparative review of the referred methods and the fifth section discusses the significance and the limitations of the findings. The sixth section concludes this study, recapitulating the key findings.

**3.2 An overview of agile methods**

In this section the existing agile methods are identified and introduced. Agile methods (shown as rectangles in Figure 1) have been characterized by the following attributes: incremental (small software releases, with rapid development cycles), cooperative (a close customer and developer interaction), straightforward (the method itself is easy to learn and to modify and it is sufficiently documented), and adaptive (the ability to make and react to last moment changes) (Abrahamsson et al. 2002). In addition to showing the various agile methods and their interrelationships, Figure 1 also presents the intellectual origins of these agile methods. In other words, these earlier studies have influenced the existing agile methods, or the ideas presented have later been used by agile method developers to build these methods. These lines of influence are represented by arrow lines. These earlier studies intellectually encroaching the existing agile methods are not, however, in the focus of this analysis. The dashed line illustrates which methods (or method developers) contributed to the publication of the agile manifesto. (http://www.agilemanifesto.org). In the following, the objectives of each method are briefly introduced.

*Adaptive software development.* Adaptive software development (ASD) (Highsmith 2000) attempts to bring about a new way of seeing software development in an organization, promoting an adaptive paradigm. It offers solutions for the development of large and complex systems, in particular. The method encourages incremental and iterative development, with constant prototyping. One ancestor of ASD is "RADical Software Development" (Bayer and Highsmith 1994). ASD claims to provide a framework with enough guidance to prevent projects from falling into chaos, but not too much, which could suppress emergence and creativity.

*Agile modeling.* Agile modeling (AM) (Ambler 2002) aims to apply the idea of agile, rapid development to modeling. The key focus in AM, therefore, is on modeling practices and cultural issues imposing values required for the application of AM. The underlying idea is to encourage developers to produce sufficiently advanced models to support acute design needs and documentation purposes. At the same time, however, AM is trying to keep the amount of models and documentation as low as possible. Cultural issues are addressed by presenting various ways to encourage communication, and to organize team structures and ways of working.

*Agile software process model.* The Agile software process (ASP) model (Aoyama 1997; Aoyama 1998) aims at enabling an accelerated development of software while maintaining flexibility to address the changing re-quirements. Aoyama and his colleagues developed ASP for the purposes of the Fujitsu company. An ancestor to ASP is the concurrent-development process (CPD) model (Aoyama 1987; Aoyama 1993). When CPD was first introduced, it represented a new emerging paradigm in software engineering (Agresti 1986). The philosophy underlying CPD was based on the principles of continuous improvement as found in Japanese production systems (Ohno 1988) for hardware assembly. ASP places emphasis on rapid and flexible adaptation to changes in process, product and environment, and it is characterized by three core processes: incremental and evolutionary process, modular and lean process, and time-based process. In Fujitsu, where ASP has been used for the development of large-scale communication systems, ASP is operationalized through the use of a network-centric tool family designed for the agile software engineering environment.

*Crystal family*. The Crystal family of methodologies (Cockburn 1998; Cockburn 2000b; Cockburn 2002) includes a number of different methods from which to select the most suitable one for each individual project. Besides the methods, the Crystal approach also includes rules of thumb for tailoring these methods to fit the varying circumstances of different projects. Each member of the Crystal family is marked with a specific color indicating the relative weight of the method. Crystal suggests choosing an appropriate-colored method for a project based on its size and criticality. Larger projects are likely to ask for more coordination and heavier or more formal methods than smaller ones. Crystal methods are open for any development practices, tools or work products, thus allowing the integration of, for example, Extreme Programming and Scrum practices.

*Dynamic systems development method.* Dynamic systems development method (DSDM) (DSDM Consortium 1997; Stapleton 1997) is a method developed by a dedicated consortium in the UK. DSDM aims at providing a control framework for Rapid Application Development. The fundamental idea behind DSDM is that instead of fixing the amount of functionality in a product, and then adjusting time and resources to reach that functionality, it is preferred to fix time and resources, and then to adjust the amount of functionality accordingly. DSDM can be seen as the first truly agile software development method.

*Extreme programming*. Extreme programming (XP) (Beck 1999; Beck 2000) is a collection of well-known software engineering practices. XP aims at enabling successful software development despite vague or constantly changing software requirements. The novelty of XP is based on the way individual practices are collected and lined up to function with each other. Some of the main characteristics of XP are short iterations with small releases and rapid feedback, close customer participation, constant communication and coordination, continuous refactoring, continuous integration and testing, collective code ownership, and pair programming. An updated version of Extreme Programming was published in 2004 (Beck and Anders 2004). The revised version divides the practices in two sets:

primary practices and corollary practices. The principle ideas remain the same nevertheless. Therefore, in our analysis we are placing the focus on the XP version 1.

*Feature-driven development.* Feature-driven development (FDD) (Coad et al. 2000; Palmer and Felsing 2002) is a process-oriented software development method for developing business critical systems. The FDD approach focuses on the design and building phases. The FDD approach embodies iterative development with the practices believed to be effective in industry. FDD emphasizes quality aspects throughout the process and includes frequent and tangible deliveries, along with accurate monitoring of the progress of the project.

*Internet-speed development.* Internet-speed development (ISD) (Cusumano and Yoffie 1999; Baskerville et al. 2001; Baskerville and Pries-Heje 2001) refers to a development situation where software needs to be released fast, thereby requiring short development cycles. ISD encompasses a descriptive, management-oriented framework for addressing the problem of handling fast releases, and includes three streams of work. In fact, the studies under ISD are considered as more management and business-oriented than other related approaches. As the first stream, Baskerville *et al.* saw that the successful management activities related to agile development framework consist of time drivers, quality dependencies and process adjustments. These are the rules under which companies have survived in ISD. In this context 'process adjustment' means focusing on good people instead of process, i.e., "if people are mature and talented, there is less need for process" (Baskerville et al. 2001, p. 56) Cusumano and Yoffie's (1999) approach to ISD draws from the "*Synch-and-stabilize*" approach by Microsoft, aimed at coping with a fast-moving, or even chaotic, software development business (Cusumano and Selby 1997). The third stream of ISD stems from the viewpoint of an emergent organization. Emergent organizations are organizations having a fast pace of organizational change – thus an opposite to stable organizations, and therefore emergent organizations are argued to need agile development tradition to survive (Truex et al. 1999). The theoretical background of the latter stream stems from Amethodological information systems (IS) development (Baskerville et al. 1992; Truex et al. 2001), which is based on a relativistic philosophy. Amethodological scholars argue that software development is a collection of random, opportunistic processes driven by accident. These processes are simultaneous, overlapping and there are gaps and the development itself occurs in completely unique and idiographic forms. Finally, the development is negotiated, compromised and capricious as opposed to predefined, planned and mutually agreed.

*Pragmatic programming.* Pragmatic programming (PP) (Hunt 2000) introduces a set of programming "best practices". It puts forward techniques that concretely augment the practices discussed in connection with the other agile methods. PP covers most programming practicalities. The

"method" itself comprises a collection of short tips (n=70) that focus on day-to-day problems. These practices take a pragmatic perspective and place focus on incremental, iterative development, rigorous testing and user-centered design.

*Scrum.* The Scrum (Schwaber 1995; Schwaber and Beedle 2002) approach has been developed for managing the software development process in a volatile environment. It is based on flexibility, adaptability and productivity. Scrum leaves open for the developers to choose the specific software development techniques, methods, and practices for the implementation process. It involves frequent management activities aiming at consistently identifying any deficiencies or impediments in the development process as well as the in the practices that are used.

*Other methods.* There are several other methods published that claim to be aligned with the agile principles as well. These include lean software development (Poppendieck and Poppendieck 2003), Evo (A method developed by Tom Gilb and summarized in Larman 2004), further developments of Rational Unified Process such as EssUP by Ivar Jacobsen and OpenUP by the Eclipse community among few others. These are left out from the analysis either because they are proprietary (EssUP) or there is a limited amount of published material otherwise (OpenUP) or they were added to the agile family of methods post-term (Evo). Lean software development's intellectual origins are connected to Toyota Production System and the manufacturing industry. Due to the focusing issues, we excluded lean thinking altogether from the analysis even if similarities may exist.

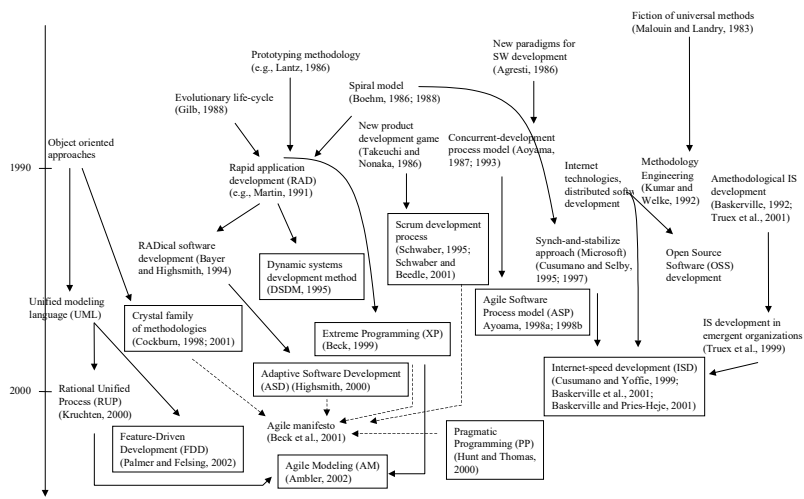

**Fig. 1. Evolutionary map of agile software development methods**.

### 3.4 Comparative review of the existing agile methods

In this section, the existing agile methods are compared using six different analytical perspectives.

#### 3.4.1 Analytical perspectives for the analysis

In order to make sense and scrutinize the existing agile methods, proper analytic tools are needed. The task of comparing any methodology with another is difficult and the result is often based upon the subjective judgment of the authors (Song and Osterweil 1991). Two alternative approaches have been proposed: informal and quasi-formal comparison (Song and Osterweil 1992). Informal comparison indicates a lack of systematic framework to guide the analysis. Quasi-formal comparison attempts to overcome the subjective limitations of the informal comparison technique by offering different strategies for composing the baseline for comparison, i.e. the analytical framework. These strategies involve the following operations (Sol 1983): 1) define an idealized method, 2) select important perspectives or features from each method, 3) derive a framework from empirical evidence, 4) use a defined metalanguage to describe each method and 5) use a contingency approach and relate the features of each method to specific problems. Song and Osterweil (1992) maintain that the second and fourth strategies are the most effective.

**Table 1. Perspectives for the analysis.**

| Perspective | Description | Key references |
|---|---|---|
| Project management support | Does the method support project management activities? | (Gilb 1988; Kumar and Welke 1992) |
| Software development life-cycle | Which stages of the software development life-cycle does the method cover? | [(Boehm 1988; Cugola and Ghezzi 1998) |
| Availability of concrete guidance for application | Does the method mainly rely on abstract principles or does it provide concrete guidance: abstract principles vs. con- | (Boehm 1988; Nandhakumar and Avison 1999) |

| Perspective | Description | Key references |
|---|---|---|
| | crete guidance? | |
| Adaptability in actual use | Is the method argued to fit per se in all agile development situations: universally predefined vs. situation appropriate? | (Malouin and Landry 1983; Kumar and Welke 1992; Truex et al. 2001) |
| Research objective | What is the method developers' research objective: critical, interpretative, means-end oriented? | (Habermas 1984; Chua 1986) |
| Empirical evidence | Does the method have empirical support for its claims? | (Kumar and Welke 1992; Basili and Lanubile 1999; Fenton 2001) |

While many analytical tools have been proposed and used to carry out the quasi-formal comparison (e.g., Olle et al. 1982; Lyytinen 1991; Hirscheim et al. 1995; Hirscheim et al. 1996; Iivari and Hirscheim 1996), The following six analytical perspectives were seen as relevant and complementary to the research purposes of the chapter. They are depicted in Table 1 with respective questions for descriptive understanding and the relevant literature references. To further exemplify these analytical perspectives, we analyze each perspective in relation to agile methods.

Methods should be efficient (as opposed to time and resource consuming) (Kumar and Welke 1992). Efficiency requires the existence of *project management* activities to enable an appropriate organization and execution of software development tasks.

A *software development life-cycle* is a sequence of processes that an organization employs to conceive, design, and commercialize a software product (Boehm 1988; Cugola and Ghezzi 1998). The software development life-cycle perspective is needed to observe which phases of the software development process are covered by the agile methods under scrutiny.

Software development methods are often used for other purposes than originally intended by their authors (Nandhakumar and Avison 1999). For example, Wastel (1996) observes that methodologies act as a social defense operating as a set of organizational rituals. Boehm (1988) concludes that the lack of concrete guidance has caused many software projects to fail, due to the fact that they have pursued development and evolution phases in the wrong order. Thus, in order to evaluate how well the methods can be used for the purposes that they have been designed for, a per-spective provided by the *availability of concrete guidance for application* is needed.

The perspective of *adaptability in actual use* stems from the works of Kumar and Welke (1992), Malouin and Landry (1983), and Truex *et al.*(2001). This perspective is used for exploring how well agile methods recognize that one ready-made solution does not fit all agile software development situations (i.e., adaptation is allowed), and if guidance is given on how to adapt an agile method in different situations (i.e., adaptation is enabled). A distinction between universally predefined and situation appropriate[2] will be drawn (Kumar and Welke 1992). Adaptability in actual use is particularly relevant in the case of agile methods, since the goal of agile software development is to increase the ability to react and respond to changing business, customer and technological needs at all organizational levels

The analysis of the agile methods in the light of their *research objectives* is used for highlighting the goals of method developers. According to Chua (1986) and Habermas (1984), research objectives include: 1) means-end oriented (technical), 2) interpretative (hermeneutical), and 3) critical (emancipatory). The means-end oriented view holds that the aim of research is to produce knowledge in order to achieve certain concrete goals or ends (Chua 1986) and to increase human control over phenomena or nature (Habermas 1984). Natural science and computer science research are typically means-end oriented, while we can also find means-end oriented research in the social or IS sciences. The goal of critical, or emancipatory, research is to point out the weaknesses of existing theories and practices – particularly dominant ones (Chua 1986).

*Empirical support* is needed for finding out what kind of empirical evidence the agile methods are grounded upon. It is often noted that while there is a great deal of anecdotal evidence to support the use of agile methods (Lindvall et al. 2002; Melnik et al. 2002), a lot less is known about the empirical evidence obtained through the use of rigorous scientific methods.

Figure 2 presents the evaluation for the first three perspectives. Each method is divided into three bars. The uppermost bar indicates the support provided for project management (analyzed in section 4.1). The middle bar indicates how well the software production process is described (pertaining to software development life-cycle analysis). The length of the bar shows which phases of software development are supported by each agile method (analyzed in section 4.2). Finally, the lowest bar shows if a method re-lies mainly on abstract principles (white color) or if it provides concrete guidance (gray color). This will be analyzed in section 4.3.

---

[2] Situation appropriate is also known in IS literature as situational awareness (Tolvanen 1998)

In general, gray color in a block indicates that the method covers the perspective analyzed while white indicates lack of such support.

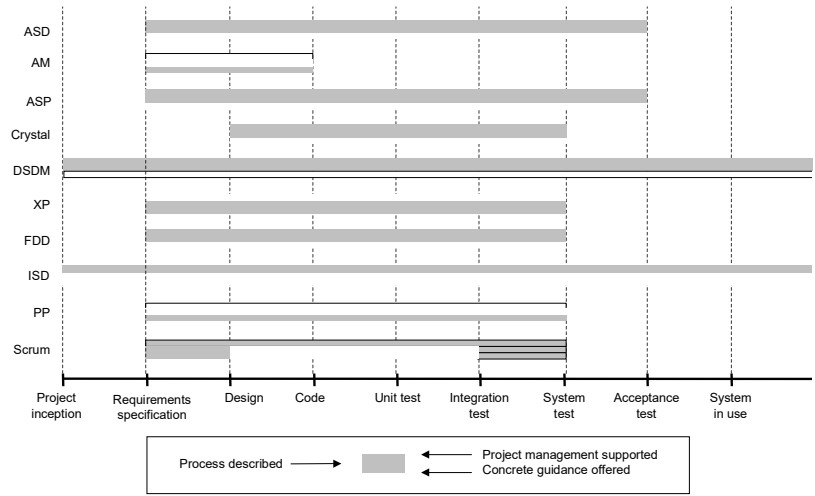

**Fig. 2.** Comparing project management, life-cycle, and concrete guidance support.

### 3.4.2 Project management

Agile software development methods differ to a large degree in the way they cover project management (uppermost bar in Figure 2). Currently, AM and PP do not address the managerial perspective at all. XP has been supplemented with some guidelines on project management (Beck and Fowler 2001), but it still does not offer a clear project management view. While some of the existing XP practices – e.g. planning game and small releases – facilitate and feed information to project management activities, the focus is on a single project and on its development. In XP the foundation for efficient project management stems from understanding the flow of user stories (i.e. requirements) from initial ideas to final product. Still, the very concept of user stories remains unclear. Beck and Fowler (2001, p. 46) explain that "[a] user story is nothing more than an agreement that the customer and developers will talk together about a feature". However, in his initial work, Beck (Beck 2000, p. 90) states that "stories are written […] with a name and short paragraph describing the purpose of the story." The basic concepts in XP, therefore, remain without a clear definition.

Scrum, on the other hand, is explicitly intended for managing agile software development projects. Thus, Schwaber and Beedle (2002) suggest the use of other methods to complement a Scrum based software development approach, naming XP as one alternative.

The approach promoted by ASD is the "adaptive (leadership-collaboration) management model" (Highsmith 2000). The focus of ASD is on changing the software development culture, and essentially, the management also having to bend in response to changes in projects.

ASP focuses on segmenting the software process so that the teams in different organizational locations can work on different software releases concurrently. ASP has a defined management hierarchy (Aoyama 1998) that displays how the work is organized at different organizational levels. ASP thus incorporates a management and organization oriented framework for large-scale agile software development.

FDD offers means for planning projects by product features, and for tracking the progress of the project. Similarly to Scrum and its management, FDD also places emphasis on empowering project managers; within an FDD project, the project manager has the ultimate say on project scope, schedule, and staffing (Palmer and Felsing 2002).

DSDM suggests a framework of controls to supplement the rapid application development approach. All of these controls are designed to increase the organizational ability to react to business changes (Stapleton 1997), which has become commonplace nowadays in all agile software development approaches. Therefore, the DSDM approach towards project management is largely about facilitating the work of the development teams, with daily tracking of the project's progress. Finally, Crystal's solution to project management focuses on increasing the ability to choose the correct method for each individual purpose (Cockburn 2002).

### 3.4.3 Software development life-cycle

In Figure 2, the overall length of the bar demonstrates which software development life-cycle phases are supported by the different agile methods. Figure 2 shows that agile methods are focused on different aspects of the software development life-cycle. DSDM is an independent method in the sense that it attempts to provide complete support over all life-cycle phases. Similarly Internet-speed development approaches also address all the phases of the software development life-cycle, but only from management and organization perspective. The others are more focused. ASD and ASP cover all the phases expect for project inception, acceptance test and system in use. AM aims at providing modeling support for requirements specification and design phases.

The Crystal family covers the phases from design to integration test. XP, PP, FDD and Scrum are focused on requirements specification, design, implementation (except for Scrum) and testing up until the system test.

From the process perspective, AM, ISD, Scrum (for the implementation part) and PP do not emphasize (or have not described) the process through which the software development proceeds. AM and PP are supplements to other methods. Thus, in the case of AM and PP the lack of process perspective seems reasonable. However, the ISD approach lacks clarity in this regard. In the other agile software development methods (i.e. ASD, ASP, Crystal, DSDM, XP, FDD and Scrum), the development process has been described.

### 3.4.3 Availability of concrete guidance for application

The lowest bar in Figure 2 indicates whether the method relies on concrete guidance (gray color) or on abstract principles (white color).

FDD lays down eight practices that "must" be used if compliance with the FDD development rules is to be valued (Palmer and Felsing 2002). The authors state that the "team is allowed to adapt them according to their experience level". Especially, in the case of these "must" practices, concrete guidance should be provided on how, in practice, this adaptation can be executed. If such guidance is missing (as it is, in this case), it is interpreted in our evaluation as reliance to abstract principles.

Based on this distinction, it was found that six out of ten agile software development methods included in the analysis placed emphasis on abstract principles over concrete guidance. ASD, again, is more about concepts and leadership-collaboration management culture than software practice. ASP (Aoyama 1998, pp. 59-65) provides the design rationale for agile software engineering environment, which describes to some extent how ASP is operationalized in Fujitsu, including the principal lessons-learned. However, ASP is a tool-centric solution for distributed agile software development. Hence, high level principles in guiding how to implement the ASP method can not be considered sufficient for other practitioners if they wish to adopt a similar working environment.

Crystal, depending on the system criticality and project size, mandates certain practices but provides little guidance on how to execute them. DSDM states that due to the fact that each organization is different no practices are detailed (Stapleton 1997). Instead, organizations should develop their practices themselves. Concrete guidance on how this should be done is not provided. Internet-speed development approaches establish certain practices or principles that should be in place. However, they do not offer any concrete guidance on how one should actually carry out the ideas of, e.g. "always analysis" or "dynamic requirements negotiation".

AM, XP and PP have been directly derived from practical settings. Their purpose and goal is to feed the collected "best practices" back into the actual practice of software development. Scrum defines the practices and offers guidance for the requirements specification phase and the integration testing phase. Implementation phases, as stated earlier, are not a part of the method.

### 3.4.5 Availability of concrete guidance for application

Majority of agile software methods allow adaptability in actual use but refrain from offering guidance on how to perform the adaptation (Table 2).

**Table 2. Adaptability in actual use**

| Perspective: | Universally predefined | Situation appropriateness | |
|---|---|---|---|
| | | Allow adaptation | Enable adaptation |
| Adaptive software development | | X | - |
| Agile modeling | | X | X |
| Agile software process | | X | - |
| Crystal family of methodologies | X | | - |
| Dynamic Systems Development Model | | X | - |
| Extreme Programming | | X | - |
| Feature-Driven Development | X | | - |
| Internet Speed development | | X | - |
| Pragmatic programming | | X | X |
| Scrum | | X | - |

Cockburn's (2002) Crystal family of methodologies explicitly provides criteria on how to select the methodology for a project. The selection is made based on project size, criticality and priority (Cockburn 2000a). On the basis of these factors a decision is made about which methodology should be used. However, when the project is underway the situation appropriateness is decreased. This interpretation is based on the fact that

Crystal methods offer prescriptive guidance. This means that Crystal, e.g., enforces certain rules such as "the projects always use incremental development cycles with a maximum increment length of four months". Thus, the adjustment is made by choosing one of the several "universal solutions", such as Crystal Clear.

FDD has goals that are grounded on imperatives or prescriptive guidance similar to Crystal. Palmer and Felsing (Palmer and Felsing 2002, p. 35) explain that FDD is "built around a core set of 'best practices'." All of these practices must be used in order to "get the full benefit that occurs by using the whole FDD process". FDD is claimed to suit "any software development organization that needs to deliver quality, business-critical software systems on time." (Palmer and Felsing 2002, p. xxiii). Thus, FDD and Crystal represent universal prescriptions that claim to have the suitability for all agile software development situations, scopes and projects.

The DSDM Consortium (1997) has published a method suitability filter in which three areas are covered: business, systems and technical (Stapleton 1997). The filter involves a series of questions such as "Are the requirements flexible and only specified at a high level?" or "Is functionality going to be reasonably visible at the user interface?" and some rationalization about which type of answer would yield greater benefits if DSDM were to be applied. While the method filter is predominantly about deciding whether the method itself is applicable or not, a recent study (Aydin and Harmsen 2002) has explicated how DSDM was adjusted to fit the purposes of an individual project, implying the necessary situation appropriate characteristics. Still, the lack of guidance on adapting DSDM in a particular software development situation indicates that the method does not *enable* adjustments on the fly.

Further, the ASD, ASP, AM, XP, ISD, PP and Scrum approaches allow situation appropriate modifications. For example, regarding XP Beck (1999, p. 77) suggests: "If you want to try XP, […] don't try to swallow it all at once. Pick the worst problem in your current process and try solving it the XP way." Beck maintains that there is no process that fits every project as such, but rather the practices should be tailored to suit the needs of individual projects. We interpret this in such a way that the number of adjustments is not limited. One of the XP practices, namely "just rules", implies that while the rules are followed, they can be changed if a mutual understanding among the development team is achieved. This implies that XP supports situation appropriateness. Again, the lack of guidance on how to perform tailoring and adaptation during the development indicates that XP lacks the adjustment enabling characteristics.

AM and PP offer supplemental practices and concrete guidance on how and when to apply these methods in actual software development work. Authors describe the situations and rationale for applying the practices suggested but refrain from offering any prescriptive guidance.

### 3.4.6 Availability of concrete guidance for application

The principal research objective of agile software method developers is means-end oriented (Table 3). Their primary goal is to provide technical solutions for the practitioners in the field.

Adaptive software development aims at offering concrete practical guidelines for helping practitioners in agile software development problems: "the book is written for project teams [using] a high-speed, high-change project to support a critical new business initiative." (Highsmith 2002a, p. xxix). This means-end oriented research objective can be observed in several places. Highsmith (2002a, p. xxv) argues that a "goal [of the book] is to offer a series of frameworks or models to help an organization employ adaptive principles." and "a goal of Adaptive Software Development is to provide a path for organizations needing to use an adaptive approach on larger projects." These extracts are further examples of the means-end oriented research objective - to offer practical solutions for solving real-life problems. Highsmith points out two major software development traditions in the history of software development: "Bureaucracy" and "adhocracy". This historical outlook can be seen as an interpretive research objective, aimed at increasing our understanding of the principles and background of these two traditions. The author also offers critiques of these approaches (Highsmith 2002a, pp. 6-8) thus engaging in critical (emancipatory) research.

The primary research objective of agile modeling (Ambler 2002) is means-end oriented, i.e. designed to present practical guidelines and examples on how "agile modelers" should do "agile modeling". There also seems to be a further critical research objective: "Current modeling approaches can often prove dysfunctional. In the one extreme, modeling is non-existent, often resulting in significant rework when the software proves to be poorly thought through. The other extreme is when excessive models and documents are produced, which slows your development efforts down to a snail' pace. AM helps you find the modeling sweet spot, where you have modeled enough to explore and document your system effectively, but not so much that it becomes a burden that slows the project down." Given that the above evidence is understood as a critique of existing modeling approaches, Ambler (2002) engages in critical research.

Aoyama's (1997; 1998) research objective falls also into the means-end oriented category, as he "proposes a new software process model, ASP (Agile Software Process) and discusses its experience" (Aoyama 1998, p. 3). In other words, Aoyama shows how ASP can be used for developing software in an agile manner.

The research objective of Cockburn (2002) is means-end oriented, as well: "These pages [Cockburn's book] contain the guidelines I use in my use case writing and in coaching: how a person may think, what he or she might observe, to end up with a better use case and use case set". Cockburn (2002), therefore, aims at presenting the best practices for use cases. Cockburn (2002) presents practical guidelines on how to increase the level of agility in software development – a means-end oriented research objective. Cockburn (2002) also engages in interpretive research by carrying out a meta-level analysis of certain concepts. He, for example, attempts to explain the meaning of a methodology thus aiming at increasing our understanding of the basic concepts, which is a sign of interpretive research objective.

**Table 3. Primary research objectives**

| Research objective: | Means-end oriented (technical) | Interpretative (hermeneutical) | Critical (emancipatory). |
|---|---|---|---|
| Adaptive software development | X | X | X |
| Agile modeling | X | | |
| Agile software process | X | | |
| Crystal family of methodologies | X | | |
| Dynamic Systems Development Model | X | | |
| Extreme Programming | X | | |
| Feature-Driven Development | | | |
| Internet Speed development | | X | X |
| Pragmatic programming | X | | |
| Scrum | X | | X |

Stapleton (1997, p. xi) argues: "DSDM provides a framework of controls and best practice for the rapid application development […] there is no book on the market that provides practical guidance on the use of the method [DSDM] or one that provides case studies from real DSDM projects." Stapleton (1997) presents the DSDM life-cycle and a nine step process, and explains how to apply these concepts in solving practical problems. This clearly refers to a means-end oriented research objective.

Beck's (1999; 2000) research objective is both means-end oriented and critical. He puts forth a set of XP practices, and explains how to apply these in industrial settings. Beck (2000) also provides guidelines for using XP principles at different levels of abstraction.

The research objective by Palmer and Felsing (2002) is means-end oriented: "spell out day-to-day details of using FDD on a real project, giving development team leaders all the information they need to apply FDD successfully to their situations." (Palmer and Felsing 2002, p. xix) This shows that their aim is to offer practical solutions to real-life problems. Palmer and Felsing also review the problems of the so-called waterfall process, but this can not be classified as critical research, as Palmer and Felsing do not themselves explore the problems of existing methods.

Internet-speed development approaches show a degree of critical (emancipatory) research . This can be perceived from the following statement: "If emergence, rather than stability, is taken as the dominant character of organizations, at least in some periods, there is a need to radically rethink the way in which IS are developed." (Truex et al. 1999, p.118) and "[Traditional information systems development] goals were highly valued by IS managers and developers, but are inappropriate for emergent organizations". Here Truex et al. (1999) criticize the traditional text-book IS/software development practices as inadequate for agile development. Baskerville et al. (2001) engage in interpretive research, since they are aiming to reveal the agile-development practice of small companies. Truex et al. (1999, p.118) reveal their interpretive research objective by explaining the background of emergence (or agile development): "In order to understand how IT can promote organizational emergence, we need to understand some of the forces behind organizational emergence."

Hunt and Thomas' (2000) research objective is primarily means-end oriented. The aim of pragmatic programming is to provide practitioners with a knowledge to "become a better programmer" (Hunt 2000, p. xvii). The authors introduce a set of programming "best practices" that cover most programming practicalities including hints and tips on how to utilize their suggestions in real-life situations. There is also a sense of interpretative analysis to be found in Hunt and Thomas' work as they attempt to describe in practical terms what the pragmatic philosophy encompasses.

The primary research objective of Schwaber and Beedle (2002) is means-end oriented: "This is a practical book that describes the experience we have had using Scrum to build systems. In this book, we use case studies to give you a feel for Scrum-based projects and management. We then lay out the underlying practices for your use in projects." (Schwaber and Beedle 2002, p.1). As seen in these extracts, the authors aim to provide practical guidance for solving real-life problems, thus placing emphasis on their means-oriented research objective. Scrum can also be regarded as critical research as it makes an attempt to show "why current system development methodologies don't work" (Schwaber and Beedle 2002, pp. 94-100). Scrum (Schwaber and Beedle 2002, pp.105-122) also touches interpretive research in attempting to explain its theoretical underpinnings in the light of Kuhn's concept of paradigm and in terms of knowledge creation as suggested by Takeuchi and Nonaka (1986).

### 3.4.7 Empirical evidence

The development of agile software development approaches are not based on systematic research (Dybå and Dingsøyr 2008). The only meta-analysis performed up-to-date in the area is done by Dybå and Dingsøyr (2008) where the authors perform a systematic review on empirical studies of agile software development. They conclude that the field is still nascent and the quality of the research falls evidently short. In what follows, each of the agile methods and the empirical research connected to these methods are summarized briefly. The list is not exhaustive but rather an illustration of the type of research conducted.

The most scientific efforts can be identified within the realm of XP practices, i.e. pair programming (Williams et al. 2000; Nawrocki and Wojciechowski 2001; Succi et al. 2002; Heiberg et al. 2003; Janes et al. 2003; Lui and Chan 2003; Williams and Kessler 2003) and test-first approach to software development (Müller and Hagner 2002), or their combination (Rostaher and Hericko 2002; George and Williams 2003). Experiences from using XP, and its variations, can be identified in university (Müller and Tichy 2001; Nawrocki et al. 2002), research institute (Abrahamsson 2003; Wood and Kleb 2003; Sfetsos et al. 2006), and commercial settings (Anderson et al. 1998; Grenning 2001; Schuh 2001; Murru et al. 2003; Rasmusson 2003). While these studies and reports provide necessary insight into the possibilities and restrictions of extreme programming, concrete data is, however, more difficult to find. Maurer and Martel's study (Maurer and Martel 2002b; Maurer and Martel 2002a) provides some data regarding the productivity gains using XP in a web development project. Wood and Kleb (2003), Abrahamsson (2003) and Hulkko and Abrahamsson (2005) have published the empirical data from their projects, which can be seen as building up the empirical body of evidence. Based on these studies, empirical hypotheses can be drawn, which are subject to further validation. For example, pair programming has been shown to reduce up to 40-50% calendar time required for job completion (Williams et al. 2000), improving job satisfaction (Williams et al. 2000; Succi et al. 2002) and producing consistently higher quality code than solo programmers (Williams et al. 2000). As an another example, Capiluppi et al. (2007) investigates evolution patterns over two and a half years for a system de-veloped using XP. The study reports that the system shows a smooth pattern of growth overall, that (McCabe) code complexity is low, and that the relative amount of complexity control work (e.g. refactoring) is higher than in other systems authors had studied. Contradictory findings are prominent as well. George and Williams (2003) have found that test-driven development takes, on average, 16% more time for development but that the resulting code quality is better. Müller and Hagner (2002) have concluded that in general test-first does not accelerate the implementation and that the results are not more reliable. Müller and Hagner (2002) base their findings on 19 students (10 in the test group and 9 in the control group) while George and Williams' (2003) subjects were 24 professional pair programmers. Based on the systematic review of 15 PP studies, Dybå et al. (2007) presented the meta-analysis to examine the effectiveness of pair programming. The analysis is reported on different angles of PP including its effect on quality, duration, effort, task complexity and expertise. The meta-analysis concluded that the PP's effectiveness "depends"— on both the programmer's expertise and the complexity of the system and tasks to be solved. PP is better for achieving correctness on highly complex programming tasks. They might also have a time gain on simpler tasks. Furthermore, on qualitative aspects on pair programming, Sfetsos (2008) investigates developer personalities and temperaments and how they affect pair effectiveness. It concludes that pairs with heterogeneous personalities and temperaments may indicate better communication, pair performance and pair collaboration-viability.

AM, ASD, Crystal, FDD, PP and Scrum have been derived from the subjective practical experience of their authors, as opposite to rigorous research. Thus, their solutions can be seen as lacking reliable empirical support. Regarding Scrum, Schwaber and Beedle (2002, p.31) claim, e.g., that "[Scrum] practices have been established through *thousands* of Scrum projects", (italics added). None of these projects are cited, however, it invites skepticism regarding the validity of the empirical evidence. Although Sutherland (2001), who is one of the originators of the method, describes how the Scrum method has evolved over years in practical setting, the details of these settings and the data are not disclosed. More recently, however, real life empirical studies on Scrum have increased substantially (Rising and Janoff 2000; Jensen and Zilmer 2003; Dingsøyr et al. 2006; Hosbond et al. 2008; Marchenko and Abrahamsson, 2008).

Agile software process and Internet-speed development were both found to have a degree of empirical support. ASP has been used for the development of large scale communication systems in Fujitsu (Aoyama 1997; Aoyama 1998). ISD, for its part, is based on interpretive (qualitative) case studies in several companies (Cusumano and Yoffie 1999; Baskerville et al. 2001; Baskerville and Pries-Heje 2001, 2004). Baskerville et al.'s case study findings have also received further empirical support (Alatalo et al. 2002). We furthermore see that these studies are aimed at increasing our understanding of how Internet-speed companies have survived, not proposing "laws" for making successful agile software development.

DSDM has been developed by a dedicated Consortium. While the aim of the Consortium has been to develop a public-domain method (Millington and Stapleton 1995), the empirical results are not openly shared with the scientific and user communities. Stapleton (1997) claims that there exists empirical evidence in the form of reports (i.e., white papers), but these are only shared with the members of the Consortium. However, since these reports have not been made publicly available, we can only conclude that the claims of DSDM are not empirically supported. Recently, Coyle and Conboy (2009) report a case study of risk management in DSDM. However, in general the empirical evidences supporting DSDM in the agile literature are almost inexistent.

The studies substantiating empirical evidences in agile methods at meta-level have been emerging recently. Dybå and Dingsøyr (2008) present a systematic review on empirical studies of agile software development. They filter thirty three agile empirical studies for analysis, in different settings, from professional projects to university courses. Dybå and Dingsøyr (2008) identify that there is a need to increase both the number and the quality of studies on agile software development. The analysis also suggests that most of the empirical studies analyzed were using XP. Scrum, one of the most popular agile methods of recent times, and having rather very low volume of empirical studies, needs further attention.

The results of the analysis confirm the initial conjectures. Empirically validated agile software engineering studies are scarce. The existing empirical studies are mostly focused on XP and its practices as well as Internet-speed development. The other methods have received considerably less attention.

### 3.5 Discussion

We now present some of the knowledge management tools at Alpha, and divide In this section, the results of the comparative review are discussed. The purpose is to identify the principal implications for research and practice. Table 4 summarizes these implications.

Project management support: software engineering is a practice-oriented field. Yet, while most (i.e. seven out of ten) agile methods do in fact incorporate some support for project management, real support is scarce. If this perspective is considered from a method feasibility point of view, efficient project management is of the utmost importance in following agile principles, such as daily builds, short release-cycles etc. Moreover, the concepts of, e.g., release and daily builds differ from one method to another. This is likely to lead to confusion rather than clarity. It appears that method developers are aiming at a niche market by deliberately using differing terminology. Practitioners, especially project managers, find themselves in a difficult position when a decision as to the most suitable approach has to be made. The operational success of any method lies in its ability to be incorporated in the daily rhythm of the software project. On this basis, we maintain that project management considerations need to be addressed explicitly to ensure alignment between developers and those managing the project.

Software development life-cycle coverage: Different agile methods cover different phases of the software development life-cycle. One possible reason for this is that most of the methods have been developed independently by different practitioners (Highsmith 2002a; Lindvall et al. 2002). Therefore, the rationale for the method focus is currently largely lacking. This hinders the possibility of effectively comparing the existing methods when trying to determine which of them is the most suitable for a given purpose. As the rationale for the life-cycle coverage is omitted, the interfaces with the phases not covered also remain unspecified, which further makes the method more difficult to apply. It is therefore suggested that the developers of agile methods pay more attention in the future to explaining their reasons for their choice of coverage, including the interfaces with the remaining phases. Information about these interfaces should include at the very least the necessary inputs for the method as well as the outputs that are generated through the application of the method.

**Table 4. Results and implications of the study**

| Perspective | Results | Implications |
|---|---|---|
| Project management support | While most methods appear to cover project management, real support is missing. | Conceptual harmonization is needed. Project management can not be neglected. |
| Software development life-cycle coverage | Current agile methods are variously focused. Reasons or rationale for their life-cycle coverage is not provided. | Life-cycle coverage needs to be explained and interfaces with phases not covered need to be clarified. |
| Availability of concrete guidance for application | Abstract principles dominate the method literature. | Emphasis should be placed on enabling practitioners to utilize the suggestions made. |
| Adaptability in ac- | Majority of the agile | More work needs to be done |

| Perspective | Results | Implications |
| --- | --- | --- |
| tual use | methods recognize that they need to be adapted to different development situations. Yet, methods lack the mechanism to enable this adaptation. | on how to adapt or adjust agile methods in different development situations. |
| Research objective | The most common research objective is means-end oriented. | Critical studies are particularly needed. |
| Empirical evidence | Empirical evidence is very limited | More empirical, situation-specific, experimental work is needed; the results need to be publicly available. |

Availability of concrete guidance for application: Only three out of ten agile methods offer concrete guidance. Thus, the current agile method literature is dominated by abstract and generic principles that offer little concrete guidance. Apparently, the agile community is more concerned about gaining acceptance of its proposed principles than offering guidance on how to use the operative versions of these principles. Currently, concrete guidance exists mostly in methods that are very limited in their life-cycle coverage (i.e. AM) or the level of detail beyond practices (i.e. PP). More work is needed to determine how the claimed practices, activities and work products can be made to operate in different organizations and situations, so as to provide practitioners with a solid base on which to formulate their decisions.

Adaptability in actual use: Some of the known agile methods (FDD and Crystal) were found to be universally predefined. Nevertheless, the methods that recognize the fact that "one size does not fit all situations", still do not offer any guidance on how this fitting process can be achieved. This being the case, forthcoming methods and studies should pay particular attention to adaptability in actual use, and offer guidance on how these methods should be used in different agile software development situations. This also requires the ability to identify the particular situations in which these fitting or adjustment activities need to be done.

The most common research objective found is means-end oriented. Of the agile methods, only a few approaches (i.e. ASD, ISD, Scrum) also have a critical research objective. Owing to the paucity of critical studies, these are particularly needed. Critical studies help us to perceive the weaknesses in the existing, perhaps dominant, agile methods and practices, while they also underline the fact that nothing should be accepted "blindly".

Empirical evidence grounded in rigorous research is scarce. The results of the analysis confirmed the initial conjectures. Only four of the ten methods studied present some empirical support for their claims. However, some methods (e.g. XP) are increasingly producing more and more empirical studies, which is bound to make the methodological base more mature. The lack of empirical evidence results in practitioners not having reliable evidence on the real usability and the effectiveness of the different agile methods. Therefore the practitioners can not determine whether the existing methods really make sense or not, wondering if the methods describe proven wisdom or if they are merely just appealing development stories. As for researchers and method developers, the lack of empirical evidence does not help in establishing a reliable and cumulative research tradition; in fact researchers do not have access to reliable evidence on existing work. Such information is, however, vital. For example, it is essential to have the knowledge of which parts of the existing work have solid foundation, and thus in deciding the extent to which we can base our future research on the existing wisdom. Thus, more empirical work is needed on the applicability of the different agile methods in different organizations and situations. Moreover, any empirical evidence should be publicly accessible. Unfortunately, the empirical studies carried out on DSDM – if such do exist – have been made available only to the members of the development Consortium. What is especially needed is empirical work exploring the effects of particular methods, their ease of use, costs, and possible negative implications for different sizes and lines of business. Such empirical work should make use of both qualitative and quantitative research methods along with data collection devices. As shown in the analysis, there exists an empirical body of evidence on the application of certain sets of agile practices. This enables the development of empirically based hypotheses with regard to the use of, e.g., pair programming and test-driven development. However, despite a great deal of literature (e.g., Grady 1992; Kitchenham 1996; Fenton and Pfleeger 1997; Hughes 2000) dealing with software metrics, what is currently lacking is a framework for measuring and validating agile software development in practical settings. This framework should define the metrics suitable for agile software development as well as provide guidance on data collection and analysis issues.

The perspectives addressed raise an important question that needs to be answered by the agile software community. Specifically, would it be more profitable to opt for extensiveness and cover more phases, or to cover fewer and aim at greater precision. In Figure 3 depth refers to the level of detail in each phase of the life-cycle. Concrete guidance supplements the level of detail by describing how the goals are achieved. Adaptability-in-use aims to make the solutions applicable to different situations.

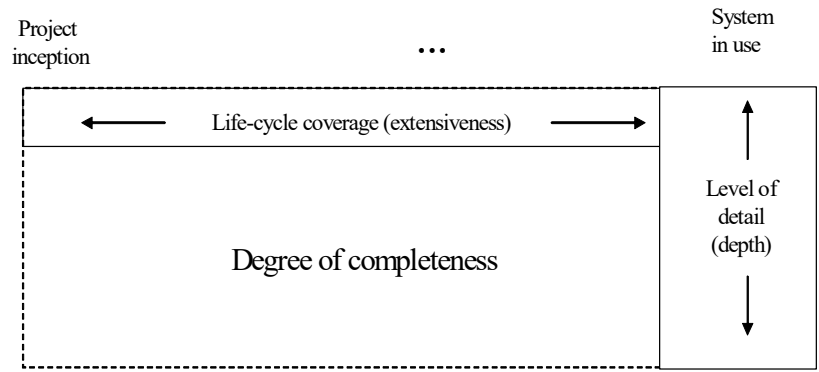

**Fig. 3. The fundamental problem of agile methods; combining extensiveness with depth in a sensible way.**

The fundamental problem here is that if a method covers all phases in great detail, it may become a methodological dinosaur, which is likely to make tailoring solutions to specific needs difficult in practice. One promising avenue of approach to this problem is method engineering (Kumar and Welke 1992), which has been applied in tailoring the DSDM in a recent case study reported by Aydin and Harmsen (2002). Completeness, a notion introduced by Kumar and Welke (1992), requires that a useful method is complete as opposed to partial. Figure 3 shows that "completeness" is an element that must be associated with both the vertical (i.e. depth of a method) and horizontal (i.e. life-cycle coverage) dimensions. None of the methods evaluated were either extensive or precise. Practitioners, currently, have only partial solutions to problems that cover a wider area than that covered by the methods. At least two possibilities can be identified. On one hand, method developers can concentrate more on specialization than generalization in their areas of expertise. Specialization refers to the development of methods or techniques to fit a particular setting or a specific development phase. An example of such a specialized approach is the Rapid7 specification technique (Kylmäkoski 2003), which focuses on efficient document authoring in software development projects. The problem here is that when methods cover too little (e.g., one phase of software development, such as requirements analysis or coding), they may become too restricted for practical purposes. Another way to approach this problem is to incorporate built-in mechanisms in to the methods to enable learning-in-use, which is likely to lead to successful method adaptation. However, methods that aim at all-inclusiveness (cover too much ground, i.e. all organizations, phases and situations) easily become too general or shallow to be useful.

### 3.6 Conclusion

The principles and methods of agile software development have provoked a substantial amount of debate and published articles. However, academic research on the subject still remains in short supply. The existing work in this field has mainly come from practicing professionals and consultants. The aim of this study has been to order and make sense of the different agile approaches that have been proposed. The existing methods were reviewed from six perspectives to determine the current situation and to suggest what directions the future agile methods should take.

In principal, it was found that the existing methods, without offering any rationale, covered various phases of the life-cycle. The majority of them do not present adequate support for project management, and only a few methods offered concrete guidance to support their solutions or adaptations in different development situations. While the method developers' research objective has been predominantly means-end oriented (technical), related empirical evidence is to date very limited.

On the basis of the above findings, new research directions were proposed. To begin with, the future agile methods need to clarify their range of applicability and to explain the interfaces to the software development life-cycle which are not part of the focus of these methods. The project management perspective cannot be neglected, either, if a method is to be welcomed in day-to-day software development practice. Emphasis should also be placed on enabling practitioners to utilize the suggestions made by the method developers. This requires more empirical work to be done on how agile methods can be adapted to different software development situations as well as studies exploring the real strengths and weaknesses of the alternative agile methods in different real-life situations. Finally, a fundamental problem was raised regarding the relationship between the coverage and depth of an agile method. If a method covers all phases in great detail, it may become a methodological dinosaur, making it difficult to apply in practice. Method engineering was proposed as a promising avenue of approach to this problem.

The current thinking on agile methods focuses on constructing a pile of conceptual methods. Instead of making haste to introduce yet more agile methods, developers should pay particular attention to the problems raised above. The field is crying out for sound methods, i.e. for methodological quality, not quantity.

**Acknowledgement**

The authors are grateful for the authors participating to the authoring of the well known VTT technical report – Agile Software Development Methods: Review and Analysis, which inspired us to take the analysis further. These are Dr. Outi Salo, Jussi Ronkainen and Dr. Juhani Warsta

**Author Biographies**

Pekka Abrahamsson is a professor in the Department of Computer Science at University of Helsinki. His research interests are centred on lean thinking, agile software development and empirical software engineering in the complex systems design space. He leads large European research projects on these topics. He is in the editorial board of Software Process Improvement and Practice and in the advisory board of IEEE Software. He is member of the ISERN, ACM and IEEE.

Nilay Oza holds a PhD is software business and he is a senior research scientist at VTT Technical research centre of Finland. He conducts re-search, develops and manages R&D projects and offers consultation to companies as a member of VTT. His current areas of research interest include agile adoption and transformation, lean thinking, Green IT business models, and global software business. He has been providing consultation in large scale agile transformation, innovation enablement, and global software business coordination. He is actively connected in international software and information systems communities.

Mikko Siponen is a Professor and Director of the IS Security Research Centre in the Department of Information Processing Science at the University of Oulu, Finland. His research interests include IS security, IS development, computer ethics, and philosophical aspects of IS. He has 33 published or forthcoming papers in journals such as *MIS Quarterly*, *Journal of the Association for Information Systems*, *European Journal of Information Systems*, *Information & Organization*, *Information Systems Journal*, *Information & Management*, *ACM Database*, *Communications of the ACM*, *IEEE Computer*, *IEEE IT Professional* and *ACM Computers & Society*. He has served as a senior and associate editor for ICIS and ECIS, and special issue SE for the MIS Quarterly. He sits on the editorial boards of the European Journal of Information Systems, Journal of Organizational and End User Computing, and Journal of Information Systems Security.